\documentclass[pra,10pt,final,twocolumn,showpacs,superscriptaddress,aps]{revtex4-1}
\usepackage{graphicx}  
\usepackage{amssymb}   
\usepackage{amsmath}   
\usepackage{color}     
\usepackage{dsfont}
\usepackage{ulem}
\usepackage{physics}
\usepackage{dsfont}

\begin{document}

\title{Dynamics of X-state in anisotropic Heisenberg spin with magnetic field}

\author{N. Vinod}
\affiliation{Department of Physics, National Institute of Technology \\ 
Tiruchirappalli - 620015, Tamil Nadu, India.}

\author{R. Muthuganesan}
\affiliation{Department of Physics, National Institute of Technology \\ 
Tiruchirappalli - 620015, Tamil Nadu, India.}

\author{R. Sankaranarayanan}
\affiliation{Department of Physics, National Institute of Technology \\ 
Tiruchirappalli - 620015, Tamil Nadu, India.}

\begin{abstract}
We study the dynamics of X-state for anisotropic Heisenberg spin system using quantum fidelity. It is shown that while Bell diagonal state, a special class of X-state, is stationary, there exists a set of two parametric states which are stationary in presence of external uniform magnetic field.
\end{abstract}

\maketitle

\section{Introduction}
Since distance measures quantify closeness of two states, they play a central role for the classification of states in quantum information theory. They are also associated with measure of geometric version of quantum correlations 
\cite{Nielsen2000}. Further, they are useful to quantify how precisely a quantum channel can transmit information. In this context, several distance measures have been introduced. Many of them were defined for classical probability distribution and then extended as distance between quantum states. Some examples of distance measures are trace distance, Hilbert Schmidt norm, Jensen-Shannon divergence and Hellinger distance \cite{Majtey2005,Bengtsson2006}. Few quantum correlation measures have also been proposed from these distance measures 
\cite{Dakic2010,Luo2011,Hu2015}.

Apart from this, there exists an important statistical measure called quantum fidelity as introduced by R. Josza. Fidelity measures the closeness between any two arbitrary mixed states $\rho$ and $\sigma$ and is defined as  $F(\rho,\sigma )= \left(\text{Tr}\sqrt{\sqrt{\rho }\sigma \sqrt{\rho }} \right)^{2}$ \cite{Jozsa1994}. Fidelity is more useful in understanding the sensitivity of perturbation in quantum evolution of a state \cite{Gorin2006}, identifying phase transition in physical systems \cite{Shi2010} and few quantum information processing such as teleportation \cite{Guo2007}, cloning \cite{Gisin1997}, tomography \cite{Bogdanov2010}. Though fidelity is not a metric, it is useful in defining distance measures such as Bures metric and sine metric \cite{metric2005}. The notion of quantum fidelity also led to other definitions of which we refer \cite{Wang2008}
\begin{align}
\mathcal{F}(\rho,\sigma) = \frac{\text{Tr}(\rho\sigma)}{\sqrt{\text{Tr}({\rho}^2)~\text{Tr}({\sigma}^2})} \label{eq:fe} 
\end{align}   
satisfying all the axioms of fidelity \cite{Jozsa1994}. For example, $\mathcal{F}(\rho,\sigma) = 1$ if $\rho = \sigma$ and $0 \leq \mathcal{F}(\rho,\sigma) \leq 1$. Further, this definition has an experimental realization using quantum networks \cite{Miszczak2009}. Recently, quantum correlations of bipartite states are defined using sine metric of the fidelity $\mathcal {F}$ 
\cite{Muthuganesan1,Muthuganesan2}. 

In this article, we investigate fidelity dynamics of X-state for a system of two spin-1/2 particles with Heisenberg anisotropic interaction in presence of an external constant magnetic field. It is shown that the fidelity dynamics is periodic for Bell diagonal state which is a special class of X-state. The periodicity is resulting to oscillation between initial state and two parametric stationary state.

\section{THE MODEL AND EVOLUTION}
The Hamiltonian of two spin-1/2 particles with anisotropic Heisenberg interaction in presence of a constant external magnetic field along $z$-axis is given as
\begin{align}
H = \frac{1}{2}\left[ J_x \, \sigma_x ^A \, \sigma_x ^B + J_y \, \sigma_y ^A \, \sigma_y ^B + J_z \, \sigma_z ^A \, \sigma_z ^B + \mathcal{B} (\sigma_z ^A + \sigma_z ^B)\right] \label{ham}
\end{align}
where $\sigma^{A(B)}_\alpha \;(\alpha= x,y,z)$ are Pauli spin matrices corresponding to particle $A(B)$, $J_\alpha$ are real coupling constants of interaction in the respective directions and 
$\mathcal{B}$ is the strength of magnetic field. By solving Sch\"ordinger equation 
$H\ket {\phi } =E\ket {\phi}$, the eigenvalues and eigenvectors of the Hamiltonian are computed as
\begin{align}
E_{1,2} =~& \frac{J_z}{2} \pm \eta, ~ \ket {\phi _{1,2}} = N_\pm \left( \frac{\mathcal{B} \pm \eta}{\Delta} \ket {00} + \ket {11}\right) \nonumber\\ 
E_{3,4} =&^{} -\frac{J_z}{2} \pm \Omega,~ \ket  {\phi _{3,4}} = \frac{1}{\sqrt{2}} (\ket {01} \pm \ket {10}) \nonumber
\end{align}
where $\eta = \sqrt{\mathcal{B}^2 + \Delta^2}$, $\Omega = (J_x+J_y)/2,~ \Delta = (J_x-J_y)/2$ and normalization constant $N_\pm=(1+\frac{\mathcal{B}\pm\eta}{\Delta})^{-1/2}$. Using the spectrum of $H$ and setting $\hbar = 1$, the time evolution operator $U(t)=e^{-iHt}$ can be written as 
\begin{align}
U(t)=\begin{pmatrix}
\mu_- & 0 & 0 & -\delta \\
0&e^{iJ_zt}\cos{{\Omega}t}&-ie^{iJ_zt}\sin{{\Omega}t}&0 \\
0&-ie^{iJ_zt}\sin{{\Omega}t}&e^{iJ_zt}\cos{{\Omega}t}&0 \\
-\delta & 0 & 0 & \mu_+
\end{pmatrix}
\end{align}
where $\mu_\pm=\cos{\eta}t \pm i\frac{B}{\eta}\sin{\eta}t$ and $\delta=i\frac{\Delta}{\eta}\sin{\eta}t$. We have omitted the overall phase factor $ e^{-itJ_z/2}$.

To investigate the dynamics of mixed state, we consider a general form of X-state as an initial state \cite{Quesada2012}: 
\begin{align}
\rho(0) = \begin{pmatrix}
	a&0&0&\omega \\
	0&b&z&0 \\
	0&z^*&c&0 \\
	\omega ^*&0&0&d \label{eq:initial}
\end{pmatrix}
\end{align}
where the diagonal entries are real and non-negative, satisfying normalization condition $\text{Tr}(\rho(0))=a+b+c+d=1$. Using local unitary transformation $I\otimes \text{diag}(e^{-i \theta _1},e^{-i \theta _2})$ or $e^{-i \theta _1 \sigma _z} \otimes e^{-i \theta _2 \sigma _z}$, it is possible to transform the X-state to that with $\omega$ and $z$ being replaced by $|\omega|$ and $|z|$, since $\theta _1$ and $\theta _2$ are arbitrary. Hence without loss of generality, we shall take the real parameters $\omega , z \geq 0$ in representing X-state. Further, non-negativity condition of density matrix implies that $z^2\leq bc$ and $\omega^2\leq ad$. The purity of the state is equal to unity if $a=d=\omega=0$ and $bc=z^2$ or $b=c=z=0$ and $ad=\omega^2$. This state generalizes many important classes of quantum states such as maximally entangled pure states (eg. Bell states), partially entangled states, quantum correlated states (eg. Werner state, isotropic state), maximally entangled mixed states, thermal state of Heisenberg two spins system etc. Entanglement and quantum correlations have been investigated for the X-state elsewhere \cite{Quesada2012,Montealegre2013,{Yurischev2015}}.
The time evolution, $\rho(t)=U(t)\rho(0)U^\dagger(t)$ provides the evolved state as
\begin{align}
\rho(t) = \begin{pmatrix}
	\rho_{11} & 0 & 0 & \rho_{14} \\
	0 & \rho_{22} & \rho_{23} & 0 \\
	0 & \rho_{32} & \rho_{33} & 0 \\
	\rho_{41} & 0 & 0 & \rho_{44}
\end{pmatrix}
\end{align}
with matrix elements
\begin{align}
\rho_{11} =& a(\mu_+\mu_-)-d(\delta^2)-\omega\delta(\mu_+ - \mu_-),\nonumber \\
\rho_{14}=& \omega(\mu_-^2 - \delta^2)+(a-d)\delta\mu_-,\nonumber \\
\rho_{22}=& b-(b-c)\sin^2\Omega t, \nonumber \\
\rho_{23}=& z-i (b-c)\sin 2\Omega t/2,\nonumber \\
\rho_{32}=&\rho_{23}^*,\nonumber \\
\rho_{33}=& c+(b-c)\sin^2\Omega t, \nonumber \\
\rho_{41}=& \omega(\mu_+^2 - \delta^2)-(a-d)\delta\mu_+,\nonumber \\
\rho_{44} =& d(\mu_+\mu_-)-a (\delta^2)+\omega\delta(\mu_+ - \mu_-).\nonumber \\
\nonumber
\end{align}
With this we obtain
\begin{align}
\text{Tr}(\rho(0))^2=\text{Tr}(\rho(t))^2=a^2+b^2+c^2+d^2+2w^2+2z^2 \nonumber
\end{align}
and 
\begin{align}
\text{Tr}(\rho(0)\rho(t))&=(a^2+d^2)(\mu_+\mu_-)+(b-c)^2(\cos^2{\Omega}t)\nonumber \\
&-2ad(\delta^2)+2bc+2z^2+2\omega^2\left(1-{\frac{2\mathcal{B}^2}{\eta^2}}sin^2{\eta}t\right). \nonumber
\end{align} 
Let us rewrite the five parametric X-state in the Bloch form,
\begin{align}
\rho(0) &=\frac{1}{4} \left[ I + s_1~\sigma _z ^A I^B + s_2~I^A \sigma _z ^B + c_1~\sigma _x ^A \sigma _x ^B \right. \nonumber \\ 
&\left. +~ c_2~\sigma _y ^A\sigma _y ^B + c_3~\sigma _z ^A\sigma _z ^B \right]
\label{is}
\end{align}
where 
\begin{align}
s_1=a+b-c-d, ~s_2=a-b+c-d, \nonumber \\
c_1=2(z+w), ~c_2=(z-w), ~c_3=a-b-c+d. \nonumber \\
\nonumber
\end{align}
In this form the coefficients are expectation values of the respective operators such that $-1 \leq s_1, s_2, c_1, c_2, c_3 \leq 1$.
With this, we have 
\begin{align}
\text{Tr}(\rho(0))^2=\text{Tr}(\rho(t))^2=\frac{1}{4}(1+s_1^2+s_2^2+c_1^2+c_2^2+c_3^2) \nonumber
\end{align}
and
\begin{widetext}
\begin{align}
\text{Tr}(\rho(0)~\rho(t)) =& \frac{1}{16}\left[2(2+2c_3^2) + 2(s_1+s_2)^2 \left( \cos^2{\eta t} + \frac{\mathcal{B}^2-\Delta^2}{\eta^2} \sin^2{\eta t}\right) - 2(s_1 - s_2)^2 +\right. \nonumber \\ 
& \left. 4(s_1 - s_2)^2 \cos^2{\Omega t} + 2(c_1 + c_2)^2 + 2(c_1 - c_2)^2 \right]- \frac{1}{4}(c_1-c_2)^2\left(\frac{\mathcal{B}^2\sin^2{\eta}t}{{\eta}^2}\right)\, . \nonumber
\end{align}
\end{widetext}

\section{BELL DIAGONAL STATE}
Here we confine our attention to three parametric Bell diagonal state, which corresponds to $s_1=s_2=0$ or $a=d$ and $b=c$, with matrix form
\begin{align}
\rho(0) = \frac{1}{4}\begin{pmatrix}
	1+c_3 & 0 & 0 & c_1-c_2 \\
	0 & 1-c_3 & c_1+c_2  & 0 \\
	0 & c_1+c_2  & 1-c_3 & 0 \\
	c_1-c_2  & 0 & 0 & 1+c_3 
\end{pmatrix}.
\end{align}
Then, the quantum fidelity between initial and evolved state state is reduced to  
\begin{align}
\mathcal{F} (\rho(0),\rho(t))=1-\frac{[c_1(0)-c_2(0)]^2~{\mathcal B}^2~\sin^2{\eta t}}{[1+c_1(0)^2+c_2(0)^2+c_3(0)^2]~{\eta}^2} 
\end{align}
where $c_i(0)$ are the Bloch coefficients of the initial state $\rho(0)$. It is interesting to observe that ${\mathcal F}=1$ for ${\mathcal B}=0$, showing that the state does not evolve in the absence of magnetic field. In other words 
\[ \frac{d\rho}{dt}=[H',\rho]=0 \,  \]
where $H'$ is the Hamiltonian (\ref{ham}) without the field.
Thus we conclude that Bell diagonal state is a stationary state of two-spin system with anisotropic Heisenberg interaction. We further note that for ${\mathcal B}\neq 0$, ${\mathcal F}=1$ if $c_1(0)=c_2(0)$. That is the Bell diagonal state with $c_1=c_2$ is a stationary state of the Heisenberg spin system even in the presence of external constant magnetic field. It is also clear that the fidelity is periodic with periodicity $T=2\pi/\sqrt{4\mathcal{B}^2+(J_x-J_y)^2}$ and for $\Delta = 0$, $T=\pi/|{\mathcal B}|$. The periodic evolution can also be written as
\[ c_1(t) - c_2(t) = [c_1(0) - c_2(0)]\cos ^2(\eta t) \, .\]   
In other words, the time evolution is periodic between initial state and the 
stationary state $(c_1=c_2)$ with period $T$. Since the sign of $c_1(t)-c_2(t)$ is the same as that of $c_1(0)-c_2(0)$, the time evolution is such that the coordinates $(c_1,c_2,c_3)$ of the evolved state do not cross the plane $c_1=c_2$.  

As an example, we consider the following initial state
\begin{align}
\rho(0)=p(\ket\phi  \bra\phi )+(1-p)\frac{I}{4}
\end{align}
with $0\leq p \leq 1$. If $p=0$, the state is a maximal mixture of orthonormal bases i.e., $\rho(0) = I/4$ with $c_i=0$, which obviously commutes with $H$ and hence is a stationary state. Further, the above state with $\ket\phi=\frac{\ket{01}\pm\ket{10}}{\sqrt{2}}$ ($c_i=p$), is also a stationary state. On the other hand, if $\ket\phi =\frac{\ket{00}\pm\ket{11}}{\sqrt{2}}$ with Bloch coefficients $c_1=c_3=-c_2=p$, the evolution is periodic such that the fidelity is
\begin{align}
\mathcal{F}(\rho(0),\rho(t))=1-\frac{4p^2~{\mathcal B}^2}{(1+3p^2)~\eta^2} \sin^2{\eta}t \, .
\end{align}
As an another example, we consider the Werner state
\begin{align}
\rho(0)=\frac{2-x}{6}I+\frac{2x-1}{6}P
\end{align}
where $P=\sum_{\alpha,\beta}\ket{\alpha}\bra{\beta}\otimes\ket{\beta}\bra{\alpha}$ 
is the flip operator with $\alpha,\beta = 0,1$ and $x\,\in [-1,1]$. Here the corresponding Bloch coefficients $c_i=(2x-1)/12$ make the state stationary.

\section{Conclusions}
In this brief article, using quantum fidelity as a tool we have investigated time evolution of five real parametric X-state under two-spin anisotropic Heisenberg interaction. The evolution of fidelity shows that while three parametric Bell diagonal state is a stationary state of the system, a two parametric subset remains stationary in presence of an external constant magnetic field. The field induces evolution of spins such that an arbitrary Bell diagonal state oscillates between initial state and the stationary state with period depends on the field strength and anisotropy.

\end{document}